# DIFFERENCE KINETIC EQUATIONS IN MANY-PARTICLE PHYSICS

A.A.Klyukanov

*Faculty of Physics, Moldova State University, 60 A. Mateevici str., Chisinau MD 2009,
Republic of Moldova,
klukanov@usm.md.*

**Abstract** – Difference Kinetic Equations are derived quantum mechanically in a plane wavelets representation with account of two-particle correlations. It is shown that the set of plane wavelet orthonormal functions $|X, K_x\rangle$ is complete. The set of ket vectors $\{|R,K\rangle\}$ is used as the second quantization basis allowing introducing the positively definite distribution function. It is obtained that inhomogeneous system is described by numbers of particles at quantized positions with quantized momenta. Difference Kinetic Equation for distribution function transforms into the classical Boltzmann Equation in the limit, where expectation value of particles number varies little in $\Delta R_i = d_i$, and $\Delta K_i = 2\pi/d_i$.

**PACS.** 71.10.-w Theories and models of many-electron systems, 71.45.-d Collective effects, 72.10. Bg General formulation of transport theory. 78.20.Bh Theory, models and numerical simulation 05.60,Gg. Quantum transport

## Introduction.

The phase space model according to which dynamical variables $r, p = \hbar k$ specify the state of a classical system gives classical description. For quantum systems a simultaneous specification of coordinates $r$ and momenta $p$ is not possible in view of Heisenberg uncertainty relations. A quantum mechanical description consists of a Hilbert space of states. According to the correspondence principle, the laws of quantum mechanics must reduce to those of classic in the limit where $\hbar$ tends to zero. This fundamental requirement views the equations of classical mechanics as limit of the Schrödinger equation. Analogously Boltzmann Equation (BE) can be derived from the Schrödinger evolution of interacting particles. The passage from a quantum description to a classical one has been explained in detail [1-3]. But classical description of homogeneous system is the same as the quantum description if one uses a plane-waves representation. Thus some of the Boltzmann Equation (BE) driving terms may be derived from the quantum mechanical many-body analysis for expectation value of microscopic polarization $P_{kk'}(t) = <a_k^+(t) a_{k'}(t)>$ by making use a plane-waves representation in a fairly direct way [1-6]. In such a manner account of particles interaction with homogeneous electric field $E$ leads to the BE drift term, represented in the form

$$\frac{\partial P_{kk'}^E(t)}{\partial t} = \frac{i}{\hbar}\sum_{k_1}[q(E \cdot r_{k'k_1})P_{kk_1}(t) - q(E \cdot r_{k_1k})P_{k_1k'}(t)] = \frac{q}{\hbar}\sum_{k_1}[P_{kk_1}(t)(E\cdot\nabla_{k_1})\delta_{k'k_1} + P_{k_1k'}(t)(E\cdot\nabla_{k_1})\delta_{kk_1}] \quad (1),$$

Here r is the radius vector, $\hbar k$ is the momentum, q is the charge of particle. Other designations are standard. The three-dimensional carrier momentum sum in Eq. (1) is explicitly evaluated analytically. Summation by part yields well known BE term

$$\frac{\partial P_{kk'}^E(t)}{\partial t} = -\frac{q}{\hbar}(E\cdot\nabla_k + \nabla_{k'})P_{kk'}, \quad \frac{\partial n_k^E(t)}{\partial t} = -\frac{q}{\hbar}(E\cdot\nabla_k)n_k, \quad n_k = P_{kk}, \quad (2)$$

Analogously BE scattering term can be easily derived in quantum theory [1-6]. But it is much more difficult to take into consideration the dependence of distribution function on position of particles which can not be considered in a plane-waves representation. In order to develop kinetic theory with simultaneous listing of coordinates and momenta one has to introduce Wigner representation [1-3, 6]. Wigner distribution function is derived from Greens function using the Wigner transform, which is Fourier transform, with respect to the relative coordinate. This technique is useful in the systems that are not homogeneous. Wigner distribution function is reduced to classic Boltzmann function in the limit where $\hbar$ tends to zero. But for many Hamiltonians of interest Wigner distribution function is not positive definite and hence can not be interpreted as a probability density. Here we develop quantum kinetic equation for microscopic polarization $P_{kk'}(t)$ that allows be considered at $k = k'$ as density of particles of inhomogeneous system described by numbers of particles at quantized positions with quantized momenta

In many-particle physics it is customary to use the term quantum Boltzmann Equation (BE) referring to the classical BE with quantum scattering integral for carrier-carrier collisions [1-4]. Thus the Boltzmann equation looks formally very similar in the classic and quantum contexts. One would think that transport theory has to be done within entirely quantum mechanical framework. But certain problems it is convenient to consider by means of distribution function $n_k$ describing the evolution of the density of particles in the classical phase space. So the advantage of the BE method is that one can use it when the system is far from the equilibrium. BE technique is also useful for study of systems that are not space homogeneous. This article represents quantum BE in the form of difference equation, which can be used for investigation and numerical simulation of quantum inhomogeneous systems.

### 1. Plane-wavelets second quantization basis set.

In terms of creation $a_k^+$ and destruction $a_k$ operators acting on Hilbert space the number of particles $n_k$ is determined as average $n_k = <a_k^+ a_k>$. Here the k subscript represents the quantum numbers set of single particle states. Angular brackets mean quantum statistic averaging with making use of the nonequilibrium density matrix. The quantum system is characterized by a projection of the many-body state on the complete set of eigenstates $\{|k\rangle\}$ of single particle Hamiltonian. In the plane-waves representation single particle state is specified by listing momentum and spin $k = k, \sigma$. Uncertainty principle states that $\Delta r_i$ tends to infinity at $\Delta k_i \to 0$. There is no Hamiltonian which quantum numbers set is exactly determined by a simultaneous enumerating of classical momenta and coordinates. Fortunately, to construct Hilbert space one only needs the complete set of functions. It is not necessary to use eigenstates of any Hamiltonian. Thus we can use the set of wavelet functions $\{|X, K_x\rangle\}$ of one-dimension system given by

$$|X, K_x\rangle = \Psi_{X, K_x}(x) = \frac{1}{\sqrt{d_x}} \exp(iK_x x) \theta_+(X + \frac{d_x}{2} - x) \theta_-(x - X + \frac{d_x}{2}), \quad K_x = \frac{2\pi}{d_x} n_x, \quad X = d_x m_x \quad (3)$$

Plane wavelet function $|X, K_x\rangle$ (3) is determined on the interval $[x - d_x/2, x + d_x/2)$. Position $X$ and momentum $K_x$ are quantized according to Eq. (3), where $n_x, m_x$-are integers $m_x = 0, \pm 1, \pm 2, \ldots \pm \left(\frac{L_x}{2d_x} - \frac{1}{2}\right)$. $n_x = 0, \pm 1, \pm 2, \ldots \pm \infty$, The number of states $(d_x dK_x / 2\pi) L_x / d_x = L_x dK_x / 2\pi$ is not dependent on number $L_x / d_x$ of discrete positions of $X$. Here $L_x$ is the length of the material system in the x direction. Asymmetrical unit step functions $\theta_+(x), \theta_-(x)$ are defined as

$$\theta_-(x) = \begin{cases} 0, & x < 0 \\ 1, & x \geq 0 \end{cases}, \quad \theta_+(x) = \begin{cases} 0, & x \leq 0 \\ 1, & x > 0 \end{cases} \quad (4)$$

Product of step functions $\theta_+ \theta_-$ in Eq.(1) is the well-known wavelet scaling function. In order to use unambiguously the set of ket vectors $\{|X, K_x\rangle\}$ as the basis of second quantization one has to show that this set is complete. The ortogonality property is expressed by means of Kroneker symbol

$$\langle X, K_x | X', K_x' \rangle = \delta_{X, X'} \frac{\sin d_x (K_x - K_x')/2}{d_x (K_x - K_x')/2} = \delta_{X, X'} \delta_{K_x, K_x'} \quad (5)$$

Equation (5) is the result of coordinates and momenta quantization (3). The closure property is proved by the analytical summation over $K_x$

$$\sum_{X, K_x} |X, K_x\rangle\langle X, K_x| = \sum_X \frac{1}{d_x} \frac{\sin\left(\pi\left(\frac{L_x}{d_x} + 1\right)\frac{x - x'}{d_x}\right)}{\sin\left(\pi \frac{x - x'}{d_x}\right)} \theta_-(x - X + \frac{d_x}{2}) \theta_+(X + \frac{d_x}{2} - x) \theta_-(x' - X + \frac{d_x}{2}) \theta_+(X + \frac{d_x}{2} - x') \quad (6)$$

Convolution with Dirichlet kernel $D(x, L_x) = \sin x(L_x / d_x + 1) / \sin x$ tends to the multiple of the delta function as $L_x \to \infty$. Thus formally one has $\delta(x) = D(x, \infty) / 2\pi$ on the interval $(-\pi, \pi)$. At

$x = m\pi, m = 0, \pm 1, \pm 2, ... \pm \frac{1}{2}(L_x/d_x - 1)$ the kernel is determined by the expression $D(m\pi, L_x) = (L_x/d_x + 1)$ but the product of step functions $\theta_+(x), \theta_-(x)$ in Eq. (6) is equal to zero at $m = \pm 1, \pm 2, ... \pm \frac{1}{2}(L_x/d_x - 1)$. Hence the closure property is expressed in the usual form

$$\sum_{X, K_x} |X, K_x\rangle\langle X, K_x| = \delta(x - x') \quad (7)$$

where $\delta(x)$ is the Dirac $\delta$-function. So we can expand the arbitrary wave function in terms of basis set $\{|X, K_x\rangle\}$ on the interval $[-\frac{L_x}{2}, \frac{L_x}{2}]$. For instance the plane-wave $|k_x\rangle = e^{ik_x x}/\sqrt{L_x}$ at $k_x = K_x$ evidently can be represented as $|k_x\rangle = \sum_X |X, K_x\rangle \sqrt{L_x/d_x}$. In the case of arbitrary $k_x$ plane wave $|k_x\rangle$ is expanded in the set $\{|X, K_x\rangle\}$ according to the equation

$$|k_x\rangle = \sum_{X, K_x} a_{X, K_x}(k_x) |X, K_x\rangle, \quad a_{X, K_x}(k_x) = \sqrt{\frac{L_x}{d_x}} \exp(i(k_x - K_x)X) \frac{\sin d_x(k_x - K_x)/2}{d_x(k_x - K_x)/2} \quad (8)$$

If the value $k_x = \frac{2\pi}{d_x} n_x$ is included into the sum over discrete set of momenta $K_x$ then we have $a_{X, K_x}(k_x) = \sqrt{d_x/L_x} \delta_{k_x, K_x}$. In the case of arbitrary $k_x \neq K_x$ the direct summation over $K_x$ in the Eq.(8) can be performed analytically

$$\frac{1}{\pi} \sum_X \sum_{n=-\infty}^{\infty} \exp\{i(k_x - \Delta K_x n)(X - x)\} \frac{\sin(k_x d_x/2)}{(k_x d_x/2\pi) - n} \theta_+(X + \frac{d_x}{2} - x)\theta_-(x - X + \frac{d_x}{2}) = 1 \quad (9)$$

to prove that the expansion of the plane wave eigenfunction $|k_x\rangle$ into the set $\{|X, K_x\rangle\}$ is given by equation (8). Expanding arbitrary function $\varphi(x)$ in terms of $\{|X, K_x\rangle\}$ using property of Dirichlet kernel and integrating over $x'$ by means of Cauchy theorem one gets

$$\int_{-\infty}^{\infty} \varphi(x') \sum_{X, K_x} \Psi^*_{X, K_x}(x) \Psi_{X, K_x}(x') dx' = \varphi(x) \quad (10)$$

If any vector can be expanded in a series of $|X, K_x\rangle$ vectors, they are said to form the complete set. Thus the set of the orthonormal functions $\{|X, K_x\rangle\}$ is complete set and can be used as the basis of second quantization. In three-dimensional systems the theory is nearby identical except there are more indices.

## 2. Problem formulation. Difference multiquantum kinetic equations.

In the $\{|R, K\rangle\}$ basis of second quantization single-particle state $k = R, K$ is characterized by the discrete set of positions $\mathbf{R}$ and momenta $\mathbf{K}$ of carriers. Hence the average $n_k = \langle \hat{a}_k^+ \hat{a}_k \rangle = P_{kk'}\delta_{kk'}$ may be interpreted as the number of the particles with momentum $\mathbf{K}$ and coordinates $\mathbf{R}$. Thus the problem reduces to determination of the expectation value of microscopic polarization $P_{kk'}(t) = \langle \hat{P}_{kk'}(t) \rangle = \langle a_k^+(t) a_{k'}(t) \rangle$. Here we develop an equation for $P_{kk'}(t)$ using multiquantum kinetic approach suggested in the reference [5]. It has been shown that motion equation for $P_{kk'}(t)$ may be represented as power series in fluctuations of the instantaneous field about the Hartree-Fock mean field. With account of pair correlations the quantum motion equation for microscopic polarization $P_{kk'}$ is represented in the form [1-8]

$$\frac{\partial \langle \hat{P}_{kk'}(t) \rangle}{\partial t} = \frac{\partial P_{k,k'}^H}{\partial t} + \left(\frac{\partial \langle \hat{P}_{kk'}(t) \rangle}{\partial t}\right)_{coll}, \quad \frac{\partial P_{k,k'}^H}{\partial t} = \frac{i}{\hbar} \langle [\hat{H}(t), \hat{P}_{kk'}(t)] \rangle \quad (11),$$

where $\hat{H}(t)$ is the one particle mean field Hamiltonian, obtained in the dynamical Hartree-Fock like approximation [1-8]. The system of kinetic equations for matrix elements $P_{kk'}(t)$ (11) is differential with respect to time and difference with respect to $k = R, K$. We consider the Hamiltonian $\hat{H}(t)$ in a simple form

$$\hat{H}(t) = \sum_{k,k'} (\hat{T} + \hat{V}(r))_{k,k'} \hat{P}_{k,k'}(t), \quad \hat{T} = \frac{\hat{p}^2}{2m}, \quad \hat{P}_{k,k'}(t) = \hat{a}_k^+(t)\hat{a}_{k'}(t), \quad \frac{\partial P_{k,k'}^H}{\partial t} = \frac{\partial P_{k,k'}^T}{\partial t} + \frac{\partial P_{k,k'}^V}{\partial t} \quad (12)$$

Here $\hat{p}$ is the momentum operator; r is the radius vector of particle. Other designations are standard. The contribution of the one particle mean field Hamiltonian $\hat{H}(t)$ to the equation of motion for microscopic polarization $P_{kk'}(t)$ (11) is divided into a drift part $\partial_t P_{kk'}^V$ describing the motion of probability distribution in $K$-space and diffusion part $\partial_t P_{kk'}^T$ describing particles motion in R-space. Motion equation in the dynamical Hartree-Fock like approximation is mathematically identical to the Schrödinger equation with mean-field Hamiltonian. The last collision term in the Eq. (11) is from the fluctuations about the mean field. BE collision term for distribution function $n_k = P_{kk}$ can be easily derived quantum mechanically [1-8]. Using results of work [5] for multiquantum Markovian processes one gets

$$\left(\frac{\partial n_k(t)}{\partial t}\right)_{coll} = \sum_{k_1}[W(k,k_1)n_{k_1}(1-n_k) - W(k_1,k)(1-n_{k_1})n_k], \quad W(k,k') = \frac{2}{\hbar}\sum_q V_q \left|e_{k,k'}^{iqr}\right|^2 \int_{-\infty}^{\infty}[n(\omega)+1]\mathrm{Im}\left\{\frac{1}{\varepsilon^*(q,\omega)}\right\}I_{k,k'}(\omega) \quad (13)$$

Here $W(k,k')$ is the transition probability, asterisk signifies the complex conjugate, $V_q = 4\pi e^2/Vq^2$, $n(\omega) = [\exp(\hbar\omega/k_0 T) - 1]^{-1}$, $\varepsilon(q,\omega)$ is the longitudinal dielectric function. In the plane-waves representation for homogeneous systems we have $e_{k,k'}^{iqr} = \delta_{k,k'+q}$. Equation (13) shows that Coulomb interaction is exposed to dynamical screening determined by the properties of dielectric function $\varepsilon(q,\omega)$ [5]. Function $I_{k,k'}(\omega)$ in Eq.(13) is given by integral

$$I_{k,k'}(\omega) = \frac{1}{\pi}\int_0^\infty \mathrm{Re}\left\{\exp\left(i\left(\int_0^t \omega_{k,k'}(s)ds - \omega t\right)\right)\right\}dt \quad (14)$$

Generalized transition frequency $\omega_{k,k'}(s)$ in the multiquantum approximation is determined by the equation (20) from reference [5]. Using quasiparticle representation we have well-known result $I_{k,k'}(\omega) = \delta(\omega_{k,k'} - \omega)$.

First of all we calculate the contribution of the mean field Hamiltonian to the equation of motion for microscopic polarization $P_{kk'}$.

### 3. V-field term $\partial_t P_{kk'}^V$. Particles motion in $K$-space.

Contribution of the potential energy $\hat{V}(r)$ to the equation of motion for $P_{kk'}$ is given by

$$\frac{\partial P_{k,k'}^V}{\partial t} = \frac{i}{\hbar}\sum_{k_1}\left(V_{k_1 k}P_{k_1,k'} - V_{k'k_1}P_{k,k_1}\right), \quad V_{k_1 k} = \delta_{R,R_1}\int_\Omega \exp(i(K-K_1)\cdot r)V(r+R)\frac{dr}{\Omega}, \quad \Omega = d_x d_y d_z \quad (15)$$

Equation (15) contains a coupling of the components of the $P_{kk'}$ matrix among themselves. It is now straightforward to insert function $V(r + R)$ into the Eq (15) and evaluate matrix elements $V_{k_1,k}$, $V_{k',k_1}$. If the function $V(r + R)$ is monotone on the interval [R-d/2, R+d/2] and varies little in d then in the dipole approximation we have

$$V(r+R) = V(R) - q(E(R)\cdot r), \quad E(R) = -\frac{1}{q}\nabla V(R) \quad (16)$$

where $E(R)$ is the inhomogeneous electric intensity. The assumption of the continuous carrier momentum leads to the replacement of the summation over $K_1$ (15) by an integral. For d insufficiently large we can not assume continuous range of value for $K_1$. In the case of discrete $K_1$ evaluating integrals $r_{k,k'}$ analytically we obtain

$$\frac{\partial P_{k,k'}^V}{\partial t} = \frac{i}{\hbar}(V(R) - V(R'))P_{k,k'}(t) - \frac{q}{\hbar}\left(\sum_{K_1 \neq K}\{(E(R) \cdot D_{K-K_1})P_{K_1R,k'}\} + \sum_{K_1 \neq K'}\{(E(R') \cdot D_{K'-K_1})P_{k,K_1R'}\}\right) (17)$$

The form of V-field term (17) is analogous to that of Boltzmann Equation; with substitution of $\nabla$ with $D$. Here $(E \cdot D) = E_i D_i$ is the scalar product. We have also introduced the notation

$$D_{K_x - K_{1x}} = \cos\left(\frac{d_x}{2}(K_x - K_{1x})\right)\delta_{K_{1y}K_y}\delta_{K_{1z}K_z}\frac{1}{K_x - K_{1x}} (18)$$

Inserting $D_{K_{1x} - K_x}$ (18) into the equation (17) one gets

$$\frac{\partial P_{k,k'}^V}{\partial t} = \frac{i}{\hbar}(V(R) - V(R'))P_{k,k'}(t) - \frac{q}{\hbar}\sum_{i=1}^{3}\sum_{n=1}^{L_i/2d_i}\frac{(-1)^n}{n\Delta K_i}\{E_i(R)(T_{\Delta K_i}^{-n} - T_{\Delta K_i}^n) + E_i(R')(T_{\Delta K_i}^{-n} - T_{\Delta K_i}^n)\}P_{k,k'} (19)$$

Shift operators with step $\Delta K_i = 2\pi / d_i$ for analytic functions are defined as

$$T_{\Delta K_i}^n = \exp(n\Delta K_i \frac{\partial}{\partial K_i}),\ T_{\Delta K_i}^{-n} = \exp(-n\Delta K_i \frac{\partial}{\partial K_i}),\ T_{\Delta K_i}^n P(K_i) = P(K_i + n\Delta K_i),\ T_{\Delta K_i}^{-n} P(K_i') = P(K_i' - n\Delta K_i) \quad (20)$$

To pass from the quantum description (19) to the classical description (2) one has to expand $P(K_x \pm n\Delta K_x)$ around $\Delta K_x = 0$ and sum the asymptotic series.

$$\frac{q}{\hbar}\sum_{i=1}^{3}\sum_{n=1}^{L_i/2d_i}\frac{(-1)^n}{n\Delta K_i}\{E_i(R)(P(K_i - n\Delta K_i) - P(K_i + n\Delta K_i))\} = -\frac{q}{\hbar}(E(R) \cdot \nabla_k)P(K) (21)$$

Hence quantum difference term $\partial_t P_{kk'}^V$ transforms to the classical in the limit where $\Delta K_x \to 0$. Contribution of the kinetic energy to the motion equation for microscopic polarization $P_{k\,k'}$ is determined analogously.

## 5. Diffusion field-term $\partial_t P_{kk'}^T$. Particles motion in the space of coordinates.

The contribution to the equation of motion for $P_{k\,k'}$ from the x-component of kinetic energy operator $\hat{T}_x = -\frac{\hbar^2}{2m}\frac{\partial^2}{\partial x^2}$ is determined by the matrix element

$$(\hat{T}_x)_{k_{1x},k_x} = \frac{\hbar^2}{2m}\int_{-\infty}^{\infty}\frac{d\Psi_{k_{1x}}^*(x)}{dx}\frac{d\Psi_{k_x}(x)}{dx}dx,\ \frac{d\Psi_{k_{1x}}^*(x)}{dx} = -iK_{1x}\Psi_{k_{1x}}^*(x) + \frac{1}{\sqrt{d_x}}\exp(iK_{1x}x)\left[\delta(x - X_1 + \frac{d_x}{2}) - \delta(X_1 + \frac{d_x}{2} - x)\right] (22)$$

The one spatial dimension carrier coordinate integral can be evaluated analytically yielding the result for T-term of difference kinetic equation in the form

$$\frac{\partial P_{k,k'}^T}{\partial t} = \frac{i\hbar}{2m}\left\{(K^2 - K'^2)P_{k,k'} + \sum_{K_1}\left([D_{R'}^2 + 2i\{(K_1 \cdot D_{R'}) - (K' \cdot D_{-R'})\}]P_{k,K_1R'} - [D_R^2 - 2i\{(K_1 \cdot D_R) - (K \cdot D_{-R})\}]P_{K_1R,k'}\right)\right\} (23)$$

Where difference operators $D$ are definite as

$$D_{R'}^2 = D_{X'}^2 + D_{Y'}^2 + D_{Z'}^2,\ D_{X'}^2 = \cos\left(\frac{d_x}{2}(K_{1x} - K_x')\right)\delta_{K_{1y}K_y'}\delta_{K_{1z}K_z'}\frac{1}{d_x^2}(T_{X'} + T_{-X'} - 2),\ (24)$$

and $\quad D_X = \cos\left(\frac{d_x}{2}(K_{1x} - K_x)\right)\delta_{K_{1y}K_y}\delta_{K_{1z}K_z}\frac{1}{2d_x}\{1 - T_X\},\ T_{X'}P(X') = P(X' + d_x) (25)$

Here one-dimensional spatial shift operator $T_X$ with step $d_x$ maps the function $P(X)$ to $P(X + d_x)$. In the limit $d_x \to 0$ at $K_1 = K, K'$ one gets

$$\frac{\partial P_{k,k'}^T}{\partial t} = \frac{i\hbar}{2m}\{(K^2 - K'^2 + \Delta' - \Delta + 2i[(K' \cdot \nabla') + (K \cdot \nabla)]\}P_{k,k'} (26)$$

This result is in accordance with classic Boltzmann Equation. If we set $k = k'$ we obtain BE term $\partial_t n_k^T = -\hbar(k \cdot \nabla)n_k / m$. At this stage we already have a closed set of difference equations. The collision term depends on the longitudinal dielectric function $\varepsilon(q, \omega)$. On the other hand, dielectric function

$\varepsilon(q,\omega)$ depends on the microscopic polarization $P_{k\,k'}$ in its turn. Hence equations (11, 12, 19, and 23) constitute a self-consistent set of difference multiquantum kinetic equations.

## 6. Difference multiquantum Boltzmann equation.

Coulomb interaction in HF like approximation [1-8] and multiquantum collision term (13) are explicitly dependent on the matrix element

$$e_{k,k'}^{iqr} = e^{iqR}\delta_{R,R'}\tilde{\delta}_{K+q,K_1}, \quad \tilde{\delta}_{K_x+q_x,K_{1x}} = \frac{Sin(d_x(K_x+q_x-K_{1x})/2)}{d_x(K_x+q_x-K_{1x})/2} \quad (27)$$

Function $\tilde{\delta}_{K_x+q_x,K_{1x}}$ differs from symbol Kroneker. At $q_x = K_{1x} - K_x$, $\tilde{\delta}_{K_x+q_x,K_{1x}} = 1$, but if $q_x \neq K_{1x} - K_x$ function $\tilde{\delta}_{K_x+q_x,K_{1x}}$ is not equal to zero, because $\Delta K_x = 2\pi/d_x$, but $\Delta q_x = 2\pi/L_x$. In the dipole approximation one has to expand $\tilde{\delta}_{K_x+q_x,K_{1x}}$ in series around $q_x = 0$. At $q_x = 0$ we have $\tilde{\delta}_{K_x+q_x,K_{1x}} = \delta_{K_x,K_{1x}}$

Using results obtained one gets difference multiquantum Boltzmann equation in the form

$$\frac{\partial n_k(t)}{\partial t} = -\frac{\hbar}{m}\sum_{K_1}((K\cdot D_{-R}) - (K_1\cdot D_R))n_{RK_1}(t) - \frac{q}{\hbar}\sum_{K_1 \neq K}(E(R)\cdot D_{K_1-K})n_{RK_1}(t) + \left(\frac{\partial n_k(t)}{\partial t}\right)_{coll} \quad (28)$$

One would think that classical BE can not be obtained from quantum BE (28) because one may not tend $\Delta K_i \to 0$ and $\Delta R_i \to 0$ simultaneously. But it is not necessary. It is instructive to note here that quantum description (28) and classical may be identical if the distribution function $n_k$ varies little in $\Delta R = d_i$, and $\Delta K_i = 2\pi/d_i$, where $\Delta R_i$ and $\Delta K_i$ don't tend to zero. Assuming that the distribution function $n_k$ varies little in $\Delta R_i$ and $\Delta K_i$ one gets the well-known differential Boltzmann equation

$$\frac{\partial n_k(t)}{\partial t} = -\frac{\hbar}{m}(K\cdot\nabla_R)n_k(t) - \frac{q}{\hbar}(E(R)\cdot\nabla_K)n_k(t) + \left(\frac{\partial n_k(t)}{\partial t}\right)_{coll} \quad (29)$$

The example of wavelet set $\{|R,K\rangle\}$ described in this article is not more than a one case where difference kinetic equation can be derived. We focus on the proof that the set of plane wavelet orthonormal functions $|R,K\rangle$ is complete. We have used the set of ket vectors $\{|R,K\rangle\}$ as the second quantization basis allowing to introduce the positively definite distribution function $n_k$. Thus the inhomogeneous system is described by numbers of particles at quantized positions with quantized momenta. Other sets of wavelet functions can be introduced if we multiply the eigenfunctions of any Hamiltonian (Bloch or Vannier functions) by scaling functions. This approach represents a major advance and provides the means for the unambiguous description and computer simulation of transport phenomena of inhomogeneous systems and nanostructures.

## References.